\documentclass[12pt]{article}
\usepackage{epsf}
\usepackage{amsmath}
\usepackage[pdftex]{graphicx}
\usepackage{color}
\usepackage{cite}
\usepackage{url}
\usepackage[pdftex,colorlinks,citecolor=blue]{hyperref}
\renewcommand{\thefootnote}{\#\arabic{footnote}}

\newcommand{\lesssim}{ \mathop{}_{\textstyle \sim}^{\textstyle <} }

\renewcommand{\thefootnote}{\fnsymbol{footnote}}
\def\thefootnote{\fnsymbol{footnote}}

\begin{document}

\begin{titlepage}

\begin{center}

\vskip .75in

{\Large \bf Cosmological bound on neutrino masses \\
in the light of $H_0$ tension} \vspace{10mm} \\ 

\vskip .75in

{\large
Toyokazu Sekiguchi$\,^a$ and  Tomo~Takahashi$\,^b$
}

\vskip 0.25in

\textit{$^{a}$ Theory Center, IPNS, KEK, Tsukuba 305-0801, Japan
\vspace{2mm} \\
$^{b}$Department of Physics, Saga University, Saga 840-8502, Japan
}

\end{center}
\vskip .5in

\begin{abstract}

Although cosmic microwave background (CMB) is the most powerful cosmological probe of neutrino masses, it is in trouble with local direct measurements of $H_0$, which is called the $H_0$ tension. Since neutrino masses are correlated with $H_0$ in CMB, one can expect the cosmological bound on neutrino masses would be much affected by the $H_0$ tension. We investigate what impact  this tension brings to cosmological bound on neutrino masses by  assuming a model with modified recombination which has been shown to resolve the tension. We argue that constraints on neutrino masses become significantly weaker in models where the $H_0$ tension can be resolved.

\end{abstract}

\end{titlepage}

\renewcommand{\thepage}{\arabic{page}}
\setcounter{page}{1}
\renewcommand{\thefootnote}{\#\arabic{footnote}}
\setcounter{footnote}{0}

\section{Introduction \label{sec:intro}}

The evidence of neutrino masses has been established by neutrino oscillation experiments which precisely measure the mass differences as  $ \Delta m_{21}^2 = (7.53 \times 0.18) \times 10^{-5}  \, {\rm eV}^2$, and $ \Delta m_{32}^2 =  (2.453 \pm 0.034) \times 10^{-3} \, {\rm eV}^2$ for normal hierarchy and  $ \Delta m_{32}^2 =  (- 2.546^{+ 0.034}_{-0.040}) \times 10^{-3} \, {\rm eV}^2$ for inverted hierarchy \cite{Zyla:2020zbs} where $\Delta m_{ij}^2 = m_i^2 - m_j^2$ with $m_i$ being the mass of the $i$-th neutrino mass eigenstate.  Since oscillation experiments cannot obtain their absolute values, other methods should pursue to probe them. Although terrestrial experiments such as tritium beta decay and neutrinoless double beta decay are such an example (see \cite{Aker:2019uuj,KamLAND-Zen:2016pfg} for recent results), indeed cosmology has been regarded as a strong tool to probe their absolute values. Among cosmological observations, cosmic microwave background (CMB) is currently the most powerful probe of neutrino masses. 

Recent data from Planck satellite provided the upper bound on the sum of neutrino masses, in combinations with other observations such as baryon acoustic oscillation (BAO), as $\sum m_\nu < 0.13 \,\, {\rm eV} \,\, (95\% \,\,{\rm C.L.})$ \cite{Aghanim:2018eyx} in the framework of $\Lambda$CDM model with neutrino masses and assuming 
degenerate mass. Bounds on neutrino mass have also been investigated
by incorporating other recent observations of large scale structure such as weak lensing \cite{Abbott:2017wau,Hikage:2018qbn}, galaxy power spectrum \cite{Cuesta:2015iho,Vagnozzi:2017ovm,Loureiro:2018pdz} and so on (see also \cite{Lattanzi:2017ubx} for a review).

When one investigates the bound on neutrino masses from CMB, one can easily notice that the neutrino masses have a degeneracy with other cosmological parameters, especially the Hubble-Lema\^itre constant $H_0$, which can also be well measured by CMB as $H_0 = (67.4 \pm 0.5)  \,{\rm km/sec/Mpc}$ \cite{Aghanim:2018eyx}. However, the value of $H_0$ obtained  by local direct measurements is $H_0=(73.8\pm1.0)\,{\rm km/sec/Mpc}$ \cite{Riess:2020sih}, which is inconsistent with the value obtained by CMB with more than 5$\sigma$ deviation. This inconsistency is now called the Hubble $(H_0)$ tension. As mentioned above, the neutrino masses have a degeneracy with $H_0$ in CMB and hence the value of $H_0$ would significantly affect the determination of neutrino masses from cosmology (see, e.g. \cite{Fukugita:2006rm,Sekiguchi:2009zs}). Therefore the $H_0$ tension is expected to give a strong impact on the cosmological bound on neutrino masses. 

The origin of the $H_0$ tension has been a target of intense research recently. It might be due to some unknown systematic errors, however it is now widely considered that the tension could indicate an extension/modification of the standard model of cosmology (see, e.g. \cite{Knox:2019rjx}, for a review). Since cosmological bounds on neutrino masses have been usually
investigated in the framework of the standard $\Lambda$CDM model\footnote{
For works in which bounds on neutrino masses have been investigated in extended models, see, e.g.,
\cite{Joudaki:2012fx,dePutter:2014hza,Chen:2016eyp,DiValentino:2016ikp,Wang:2016tsz,Canac:2016smv,Bellomo:2016xhl,Lorenz:2017fgo,Vagnozzi:2018jhn,Zhao:2018fjj,Poulin:2018zxs,Zhang:2020mox}. 
}, if the tension is resolved by extending/modifying the cosmological model, the neutrino masses should be re-investigated in such a new framework since the bound might be significantly affected, which is the issue we would like to argue in this paper\footnote{
We in this paper consistently assume that neutrinos are the standard active ones and the sum of the masses is the only relevant parameter associated to them.
}.

Although there is no consensus on the plausible model to solve the $H_0$ tension, the present authors have recently proposed a model which can significantly resolve the tension, based on a model with early recombination \cite{Sekiguchi:2020teg}. Since this model satisfies the necessary conditions which a plausible scenario should share to solve the $H_0$ tension \cite{Sekiguchi:2020teg}, a model with early recombination can be regarded as a representative one as a solution for the tension. Therefore we adopt this model and investigate a constraint on neutrino masses in the model by using cosmological observations such as CMB and BAO and compare its constraint obtained in the framework of $\Lambda$CDM,  which would highlight the impact of the $H_0$ tension to the cosmological bound on neutrino masses. 

The structure of this paper is as follows. In the next section, we discuss the setup of our analysis where we briefly review a model with early recombination proposed in \cite{Sekiguchi:2020teg}  and explain our method of the analysis to investigate a cosmological bound  on neutrino masses. Then in Section~\ref{sec:results}, we show our results on constraints on neutrino masses in the framework of early recombination and flat and non-flat $\Lambda$CDM and then, make a comparison between those constraints. As mentioned above, the model with early recombination would have properties which a successful model for the $H_0$ tension  should share and hence the bound obtained in the framework would show general tendencies in models where the $H_0$ tension is solved.  The final section is devoted to conclusion of this paper.

\section{Setup of the analysis \label{sec:setup}}

In this section, we discuss the setup of our analysis. First we briefly review a model with early recombination which can significantly relax the $H_0$ tension and be regarded as a representative model to solve the tension. Then we summarize the  method of our analysis to constrain neutrino masses from cosmological observations such as CMB, BAO and type Ia supernovae (SNeIa). We also make an analysis including the Planck lensing data and the local $H_0$ measurements.

\subsection{Model with early recombination \label{sec:early_recomb}}

Here we briefly describe a model with early recombination we adopt in our analysis as a representative model to solve the $H_0$ tension. For the details of the model, we refer the readers to Ref.~\cite{Sekiguchi:2020teg}.

As  mentioned in the introduction, this model satisfies the necessary conditions which a successful solution would meet\footnote{
These necessary conditions are listed in \cite{Sekiguchi:2020teg}.
}.  Among the conditions,  the most non-trivial one is to reduce the sound horizon at recombination $r_s (z_\ast)$\footnote{
Precisely speaking, this should be the sound horizon at the drag epoch, however, given $r_s (z_\ast)$, the one at the drag epoch can also be determined. Therefore we use $r_s (z_\ast)$ in the following.
} with $z_\ast$  being the redshift at recombination epoch, by $\sim10$\% compared to the value obtained by fitting to Planck data in the $\Lambda$CDM model,  keeping the fit to the CMB (Planck data) remains unchanged. A model with the early recombination realizes this condition in a non-trivial way. As a possible realization of early recombination scenario, we adopt a model with time-varying electron mass~$m_e$.  Although we can explicitly show that $r_s (z_\ast)$  can be reduced by $\sim\mathcal O(10)$\% without spoiling the fit to CMB by directly calculating the CMB power spectrum numerically, we can also argue analytically to some extent by using some key quantities which characterize the CMB power spectrum.

We can approximately well describe the effect of early recombination  by the shift of the scale factor at recombination epoch $a_\ast$. The change of  the recombination epoch affects the CMB power spectrum, which is characterized by the changes of the position and height of acoustic peaks and the diffusion damping.  Regarding the height of the acoustic peaks,  the following two quantities well describe it:
\begin{eqnarray}
\label{eq:R}
R(x) &=& \displaystyle\frac{3\omega_b}{4 \omega_\gamma} = \frac{3\omega_b a_*}{4\omega_\gamma}x  \,, \\ [8pt] 
\label{eq:dtauda}
A(x) &=& a^2 H  = \frac{H_0}{h}\sqrt{\omega_ma_*x+\omega_r} \,, 
\end{eqnarray}
where we have introduced a quantity $x \equiv a / a_\ast$ with the scale factor being normalized by the one at recombination. $\omega_b \,  (= \Omega_b h^2), \omega_m \,(=\Omega_m h^2) $, $\omega_\gamma \,  (= \Omega_\gamma h^2)$ and $\omega_r \,  (= \Omega_r h^2)$ represent energy densities of baryon, total matter, photons and radiation (assuming neutrinos are sufficiently relativistic by the time of recombination) with $\Omega_i $ being the normalized energy density for a component $i$ and $h$ being the reduced Hubble constant in units of 100 km/s/Mpc (i.e., $H_0 = 100 \, h$ km/s/Mpc).  The former quantity $R(x)$, the ratio between baryon and photon densities, gives the relative height of even and odd peaks. The latter $A(x) $ characterizes the integrated Sachs-Wolfe (ISW) effect which determines the height of the first peak. From the above formulas, the shift of $a_\ast$ can leave $R$ and $A$ unchanged by changing $\omega_m$ and $\omega_b$ as 
\begin{equation}
\label{eq:Delta_b_m}
\Delta_{\omega_b} = \Delta_{\omega_m} = - \Delta_{a_\ast} \,,
\end{equation}
where $\Delta_i$ denotes a fractional change of a quantity $f$ from its reference value $\Delta = (f - f_{\rm reference})/ f_{\rm reference}$. 

The sound horizon at recombination is given by 
\begin{equation}
r_s(a_*) = \frac{a_*}{\sqrt3}\int^1_0 \frac1{\sqrt{1+R(x)}} \frac{dx}{A(x)},
\end{equation}
from which one can see that $r_s (a_\ast)$ also changes by the shift of $a_\ast$ as 
\begin{equation}
r_s (a_\ast) \propto a_\ast 
\end{equation}
when Eq.~\eqref{eq:Delta_b_m} is satisfied.
On the other hand, the diffusion damping (Silk damping) scale $1/k_D$ is given by 
\begin{equation}
\frac1{k_D(z_*)^2}=\frac{a_*^2}6 \displaystyle\int^1_0 \frac{R^2+\displaystyle\frac{16}{15}(1+R)}{(1+R)^2} 
\frac{1}{a_*^2 n_e \sigma_T} \frac{dx/x}{A} \,,
 \label{eq:defkD}
\end{equation}
where $\sigma_T$ and $n_e$ are the Thomson scattering cross section and the electron number density. 
To keep the CMB power spectrum intact, 
the ratio between $1/k_D$ and the sound horizon at recombination epoch $r_s (a_\ast)$ should be kept unchanged.
In other words,
\begin{equation}
1/k_D \propto a_\ast
\end{equation}
should be satisfied. This is satisfied in an early recombination model if
\begin{equation}
a_*^2n_e\sigma_T=x_e \frac{1-Y_p}{m_H} \frac{\rho_{\rm crit}}{h^2}(\omega_ba_*)\left(\frac{\sigma_T}{a_*^2}\right)\frac1{x^3},
\label{eq:f}
\end{equation}
is kept unchanged as a function of $x$.

Finally, we also need to keep the viewing angle of the sound horizon untouched, which is represented by the quantity $\theta_s (a_\ast) \equiv r_s (a_\ast) / D_M (a_\ast)$ with $D_M (a)$ being the angular diameter distance to $a (=1/(1+z))$:
\begin{equation}
D_M(z) = 
\begin{cases}
\displaystyle\frac {\sin\left[\sqrt{-\Omega_k}H_0\chi(z)\right]}
{\sqrt{-\Omega_k}H_0} & \mbox{for~}\Omega_k<0\mbox{~(closed)} \\ \\
\chi(z) & \mbox{for~}\Omega_k=0\mbox{~(flat)}\\  \\
\displaystyle\frac 
{\sinh\left[\sqrt{\Omega_k}H_0\chi(z)\right]}
{\sqrt{\Omega_k}H_0} & 
\mbox{for~}\Omega_k>0\mbox{~(open)}
\end{cases} \,, 
\label{eq:DM}
\end{equation} 
where $\chi$ is the comoving distance to $z$ which is given by 
\begin{equation}
\chi(z)=\int^z_0\frac{dz}{H(z)} \,.
\end{equation}
Around the mean cosmological parameters of the $\Lambda$CDM model from the Planck 2018 result \cite{Aghanim:2018eyx}, the change of $\theta_s (a_\ast)$ can be canceled by shifting the Hubble parameter, in the $\Lambda$CDM background, as 
\begin{equation}
\label{eq:Delta_h}
\Delta_h \simeq -3.23  \Delta_{a_\ast} \,.
\end{equation}
Even when we consider a different background, the relation $\Delta_h \propto - \Delta_{a_\ast}$ holds, which introduces a strong degeneracy between $H_0$ and $a_\ast$.

From the above argument, one  can see that the  reduction of the sound horizon is realized by the change of $a_\ast$ whose effects on CMB power spectrum can be canceled by changing other cosmological parameters as given in Eqs.~\eqref{eq:Delta_b_m} and \eqref{eq:Delta_h} once Eq.~\eqref{eq:f} is satisfied.
 Importantly, $H_0$ can be shifted to a higher value by taking recombination epoch earlier, which can solve the $H_0$ tension. 

As mentioned above, the early recombination can be realized by assuming a time-varying electron mass $m_e$. The effects of varying $m_e$ can be understood by noting that: (i)~$m_e$ changes the energy level of hydrogen as $E \propto m_e$, (ii)~Thomson scattering cross section is affected as $\sigma_T \propto m_e^{-2}$. These effects amounts to the shift of the recombination epoch as
\begin{equation}
\Delta_{m_e} = - \Delta_{a_\ast} \,,
\end{equation}
with Eq.~\eqref{eq:f} being kept unchanged automatically.
Therefore a model with early recombination can be realized by assuming a time-varying $m_e$  and can solve the $H_0$ tension as far as CMB power spectrum is concerned\footnote{
For the effects of time-varying electron mass on CMB, see also \cite{Hart:2017ndk,Hart:2019dxi}.
} since the fit to the CMB is automatically kept unchanged.

We can actually show that, by varying $m_e$, $H_0$
can be shifted to a higher value which can significantly relax the tension without spoiling the fit to CMB. However, when we combine the data from BAO and SNeIa, the distance measure cannot be well-fitted in the framework above and we need to modify the background evolution after recombination. This might be done in several ways, but here we consider a simple extension,  a non-flat Universe to realize this since  we just introduce one additional free parameter in this case: the curvature of the Universe $\Omega_k$. Therefore, in the following we investigate bounds on neutrino masses in a model with varying $m_e$ in a non-flat Universe. We refer this model as $m_e \Omega_k \Lambda$CDM  for brevity in the following.

\subsection{Analysis}

We investigate cosmological bound on neutrino masses from the data from Planck (TT, TE, EE + LowE) \cite{Aghanim:2019ame}, BAO \cite{Beutler:2011hx,Ross:2014qpa,Alam:2016hwk} and SNeIa \cite{Scolnic:2017caz} by performing Markov Chain Monte Carlo (MCMC) analysis. In addition, we optionally also include CMB lensing data from Planck \cite{Aghanim:2018oex} and the direct measurements of Hubble-Lema\^itre constant, $H_0\,[{\rm km/sec/Mpc}]=74.1\pm 1.3$ from \cite{Riess:2020sih}\footnote{
We adopt the results without SNeIa in order to minimize systematic errors associated to SNeIa data.
}, which we denote as H0 in the following.
In parameter estimation, we use a modified version of {\tt CosmoMC} \cite{Lewis:2002ah} which accommodates the time-varying electron mass supported by the recombination code {\tt HyRec} \cite{AliHaimoud:2010dx,hyrec}. 
We note that, although we discussed the effects of modified recombination or time-varying electron mass just focusing on the change of the recombination epoch in the previous section,  {\tt Hyrec} code adopted in the analysis incorporates its full effects. We refer the readers to Ref.~\cite{hyrec}  for detail.
 
We assume the degenerate mass hierarchy for neutrinos and investigate cosmological constraints on $\sum m_\nu$ in 
a canonical flat $\Lambda$CDM background and an early recombination model with varying $m_e$ and nonzero spatial curvature ($m_e\Omega_k\Lambda$CDM). 
For reference, we also consider a non-flat $\Lambda$CDM model ($\Omega_k\Lambda$CDM). The primary parameters in our analysis for the $\Lambda$CDM model are: cold dark matter density $\omega_c $,  baryon density $\omega_b$, the acoustic angular scale $\theta_{\rm MC}$, the reionization optical depth $\tau$, the amplitude of primordial power spectrum $A_s$, the spectral index $n_s$ and the sum of neutrino masses $\sum m_\nu$. In the analysis in the framework of the $\Omega_k\Lambda$CDM model, the curvature density $\omega_k ( = \Omega_kh^2)$ is also varied in addition to the above parameters. For the case of the $m_e\Omega_k\Lambda$CDM model, the electron mass $m_e$ is also included as a free parameter. When the electron mass is varied, we assume that $m_e$ becomes the standard value some time after recombination so that it does not affect late time Universe. Flat priors are assumed for all primary parameters in the analysis.

\section{Results \label{sec:results}}

\begin{figure}
\centering
\includegraphics[width=9cm]{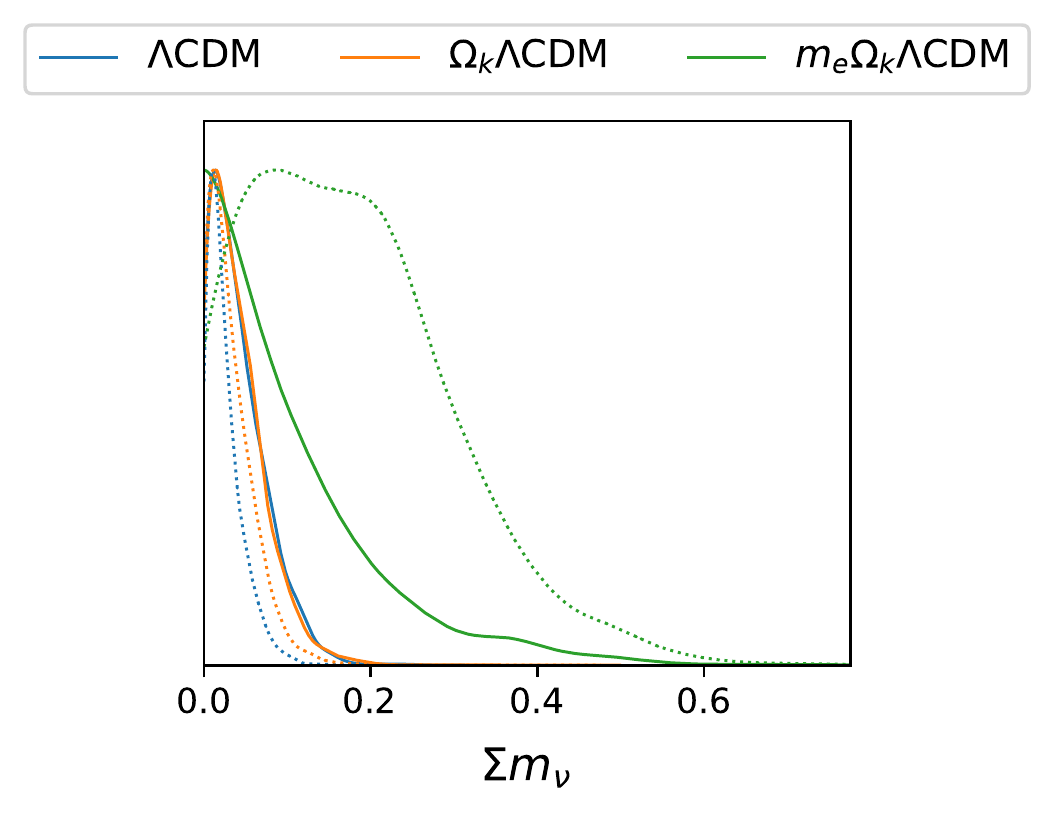}
\caption{ \label{fig:1D} 
1D posterior distributions neutrino masses from CMB+BAO+SNeIa+lensing (solid line) and  CMB+BAO+SNeIa+lensing+H0
(dotted line), where the cosmological backgrounds are assumed to be $\Lambda$CDM (blue), $\Omega_k\Lambda$CDM (orange) and $m_e\Omega_k\Lambda$CDM (green). }
\end{figure}

\begin{figure}
\centering
\includegraphics[width=16cm]{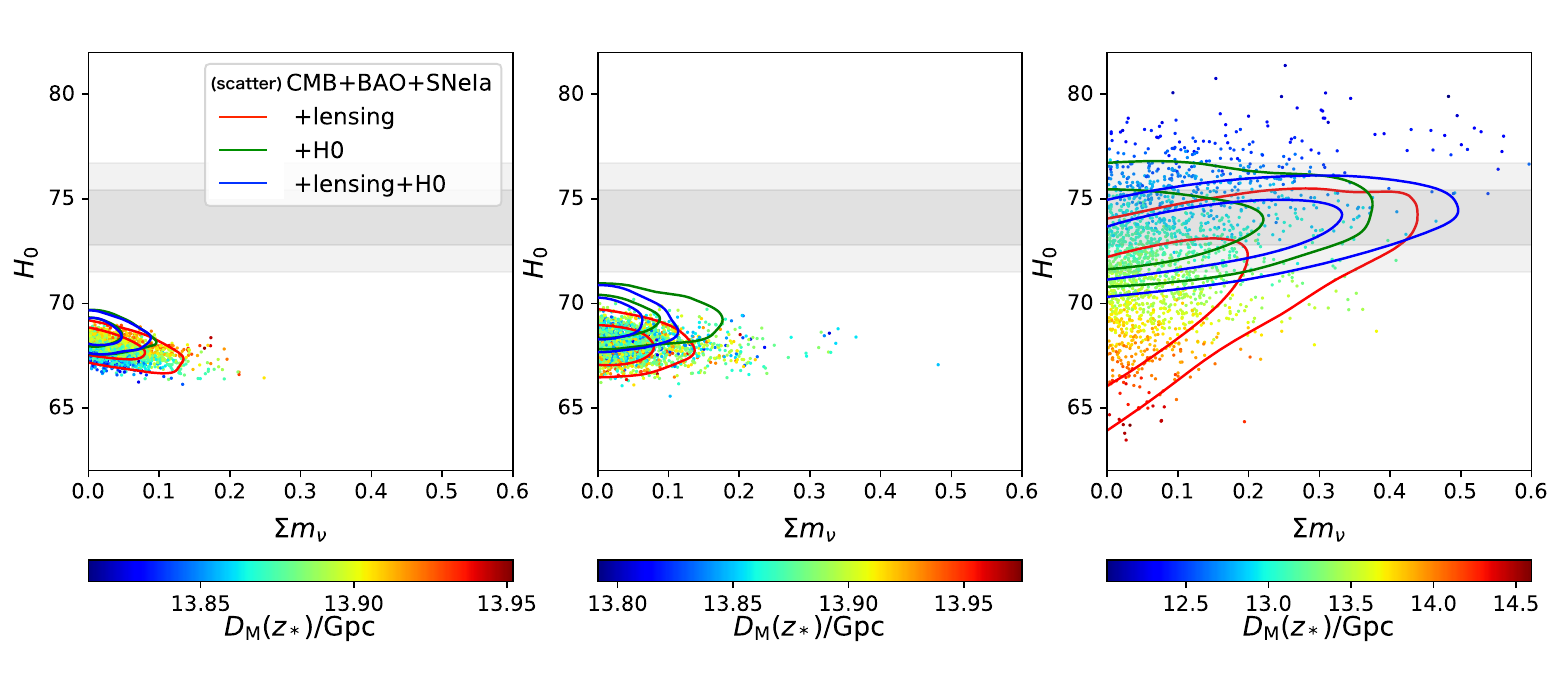}
\caption{\label{fig:2D} Constraints in the plane of neutrino masses and $H_0$ in the framework of $\Lambda$CDM (left), $\Omega_k\Lambda$CDM (middle) and  $m_e \Omega_k  \Lambda$CDM (right) models.  1$\sigma$ and 2$\sigma$ allowed regions are shown for the analysis of CMB+BAO+SNeIa+lensing, CMB+BAO+SNeIa+H0 and CMB+BAO+SNeIa+lensing+H0 are depicted. Scatter plots of $D_M(z_\ast)$ for the analysis of CMB+BAO+SNeIa are also shown. Grey horizontal shaded band indicates the values obtained from the local $H_0$ measurement. }
\end{figure}

\begin{figure}[h]
\centering
\includegraphics[width=14cm]{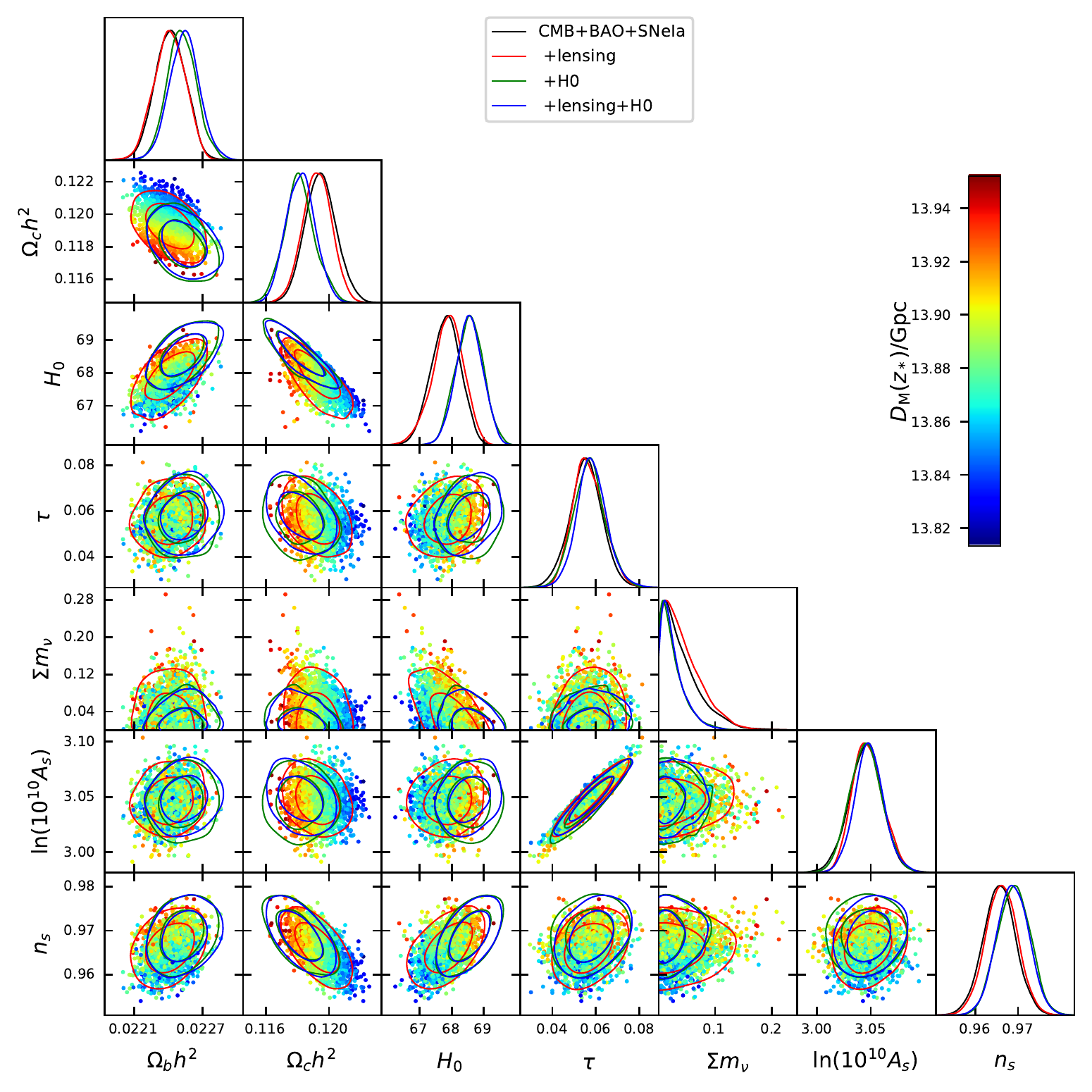}
\caption{Triangle plot of cosmological parameters in the $\Lambda$CDM model.  \label{fig:triangle_LCDM} }
\end{figure}

Now we present our results. First we show 1D posterior distribution for neutrino masses in the framework of flat and non-flat $\Lambda$CDM (i.e., $\Lambda$CDM and $\Omega_k\Lambda$CDM) models and 
the modified recombination in non-flat Universe (i.e., $m_e\Omega_k\Lambda$CDM model) in Fig.~\ref{fig:1D}.  95\% C.L. upper bounds on $\sum m_\nu$ are summarized in Table \ref{tab:mnu}.  Constraints in the $\sum m_\nu$-$H_0$ plane for some combinations of data sets and the scatter plot of the angular diameter distance to last scattering surface $D_M (z_\ast)$ for the analysis of CMB+BAO+SNe are shown in Fig.~\ref{fig:2D}.  Full triangle plots for flat $\Lambda$CDM, non-flat $\Lambda$CDM and $m_e\Omega_k\Lambda$CDM models are respectively depicted in Figs.~\ref{fig:triangle_LCDM}, \ref{fig:triangle_OLCDM} and \ref{fig:triangle_meOLCDM}, in which 1D posterior distribution and 2D allowed regions for the analysis of CMB+BAO+SNeIa+lensing, CMB+BAO+SNeIa+H0 and CMB+BAO+SNeIa+lensing+H0 are shown for the primary parameters except the acoustic angular scale $\theta_{\rm MC}$ being replaced by $H_0$. In 2D panels, scatter plots for the angular diameter distance to last scattering surface $D_M(z_\ast)$ are depicted for the analysis of CMB+BAO+SNeIa to discuss the degeneracies among the parameters.

When a flat $\Lambda$CDM is assumed, CMB+BAO+SNeIa+lensing gives $\sum m_\nu <0.11~{\rm eV}$  (95 \% C.L.), which is consistent with Planck 2018 result \cite{Aghanim:2018eyx}. As long as the Planck lensing data is  included, flat and non-flat $\Lambda$CDM background gives similar constraints as read off from Fig,~\ref{fig:1D} and Table~\ref{tab:mnu}. However, in models with modified recombination (i.e., varying $m_e$) in a non-flat framework, the upper bound is significantly weakened, which suggests that in a scenario where the $H_0$ tension can be solved, cosmological constraint on neutrino masses gets less severe.  We will take a closer look for each model below.

\begin{table}
\centering
\begin{tabular}{l|c|c|c}
\hline\hline
& $\Lambda$CDM$+m_\nu$ & $\Omega_k\Lambda$CDM$+m_\nu$ & $m_e \Omega_k\Lambda$CDM$+m_\nu$ \\
\hline
CMB+BAO+SNeIa & 0.11 & 0.16 & 0.28\\
\quad+lensing & 0.11 & 0.11 & 0.34 \\
\quad+H0 & 0.072 & 0.14 & 0.31\\
\quad+lensing+H0 & 0.069 & 0.089 & 0.40\\
\hline\hline
\end{tabular}
\caption{95\% upper bounds on $\sum m_\nu$\,[eV].\label{tab:mnu} }
\end{table}

\subsection{Case of $\Lambda$CDM model} 
In the case of $\Lambda$CDM model, when we incorporate the local $H_0$ measurements,  the upper bounds on $\sum m_\nu$ are superficially tightened. This is because the neutrino mass and $H_0$ are negatively correlated in the CMB data in the $\Lambda$CDM model \cite{Sekiguchi:2009zs}, and hence the local measurement of $H_0$, which prefers a large $H_0$, inevitably leads to a lower $\sum m_\nu$ as seen from the left panel of Fig.~\ref{fig:2D}.  The scatter plot of $D_M (z_\ast)$ in Fig.~\ref{fig:2D} also shows that the degeneracy between $\sum m_\nu$ and $H_0$ corresponds to the direction of constant $D_M (z_\ast)$. When $\sum m_\nu \lesssim 0.1$\,eV, neutrinos become non-relativistic well after the recombination, and hence, for this magnitude of $\sum m_\nu$, neutrino masses only marginally change perturbation evolution by the time of recombination. Primary effects of neutrino masses on CMB anisotropy therefore should arise from the modification to late-time expansion, namely the distance to  last scattering surface $D_M (z_\ast)$ \cite{Fukugita:2006rm}. Given the fact that $\Omega_bh^2$ and $\Omega_ch^2$ are tightly constrained by spectral shape of the CMB power spectrum, $H_0$ (or $\Omega_\Lambda h^2$) is the only cosmological parameter which can cancel the effects of neutrino masses to $D_M (z_\ast)$ in a flat $\Lambda$CDM model, as far as only CMB power spectrum is concerned. The heavier the neutrino masses get, the earlier neutrinos become non-relativistic, which makes the angular diameter distance to $z_\ast$ smaller. To keep $D_M(z_\ast)$ unchanged, $H_0$ should be taken to be smaller, which explains the negative correlation between $\sum m_\nu$ and $H_0$. Allowed region in the $\sum m_\nu$-$H_0$ plane in this model reflects this fact and are in significant tension with the local $H_0$ measurement. In other words, nonzero neutrino masses exacerbate the Hubble tension.  Even late-time distance measurements, {\it i.e.} BAO and SNeIa, lift the degeneracy only slightly.

\subsection{Case of $\Omega_k \Lambda$CDM model} 
\begin{figure}[h]
\centering
\includegraphics[width=15cm]{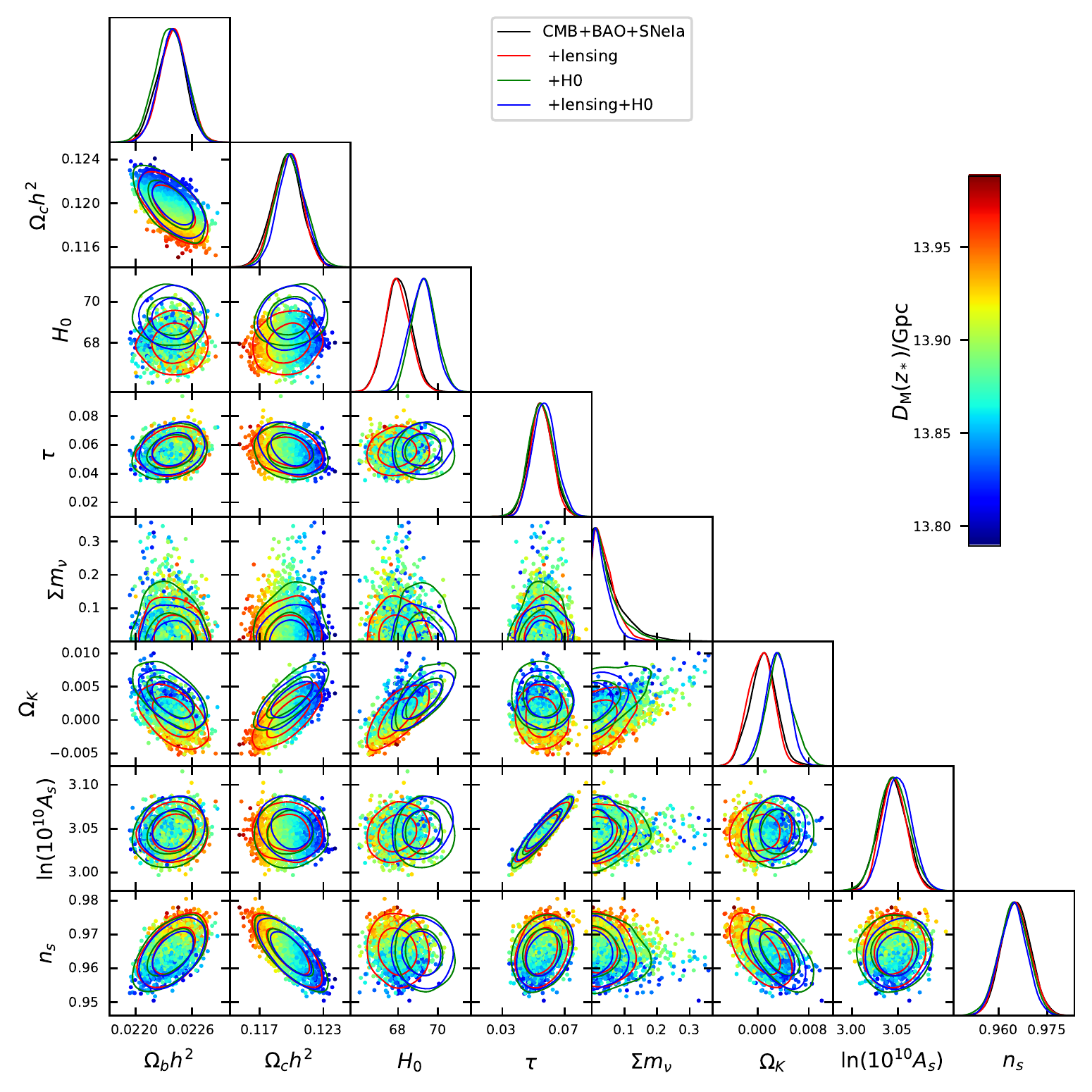}
\caption{Triangle plot of cosmological parameters in the $\Omega_k  \Lambda$CDM model.  
\label{fig:triangle_OLCDM} }
\end{figure}

In a non-flat $\Lambda$CDM model (i.e., $\Omega_k \Lambda$CDM),  the curvature of the Universe can also affect the angular diameter distance to last scattering surface $D_M(z_\ast)$. As mentioned above, $D_M(z_\ast)$ can be modified by changing  neutrino masses, however, $\Omega_k$ is more powerful in changing $D_M(z_\ast)$ than neutrino  masses, $H_0$ is mainly degenerate with $\Omega_k$, which can be read off from the panel showing the constraint in the $H_0$--$\Omega_k$ plane in Fig.~\ref{fig:triangle_OLCDM}. As seen from the figure, $H_0$ and $\Omega_k$ degenerate along a constant $D_M(z_\ast)$. On the other hand,  due to the existence of $\Omega_k$, the degeneracy between $\sum m_\nu$ and $H_0$ gets significantly weakened and almost disappears in the $\Omega_k \Lambda$CDM model as seen from the middle panel of Fig.~\ref{fig:2D}. Scatter plot of $D_M(z_\ast)$ in the panel also suggests that $D_M (z_\ast)$ is almost irrelevant to set a constraint in the $H_0$--$\sum m_\nu$ plane, which is quite different from the case of a flat $\Lambda$CDM model. 

It should be noted here that, although the degeneracy between $H_0$ and $\sum m_\nu$ disappears in the $\Omega_k \Lambda$CDM model, $\Omega_k$ and $\sum m_\nu$ are degenerate along the direction of a constant $D_M(z_\ast)$, which makes an upper bound on neutrino masses weaker. When the local $H_0$ measurement is included, the degeneracy between $H_0$ and $\Omega_k$ is broken, which in turn makes an upper bound on neutrino masses slightly severe. 
However, it should be noted that, due to the indirect effect of the degeneracy between $H_0$ and $\Omega_k$,  constraints on neutrino masses get weaker compared to the one in the $\Lambda$CDM model.
In any case, the inclusion of neutrino masses does not improve the $H_0$ tension in the $\Omega_k \Lambda$CDM model as well.

\subsection{Case of $m_e \Omega_k \Lambda$CDM model} 
Finally, we discuss the case of  the $m_e\Omega_k\Lambda$CDM model which has been suggested as a solution to the $H_0$ tension  \cite{Sekiguchi:2020teg}. As seen from Fig.~\ref{fig:1D} and Table~\ref{tab:mnu}, when we assume the $m_e\Omega_k\Lambda$CDM model, the constraint on $\sum m_\nu$ is relaxed significantly from the ones for flat and non-flat $\Lambda$CDM models. In models with varying $m_e$,   the recombination epoch can be altered, which substantially changes  $r_s (z_\ast)$ as we have discussed in Section \ref{sec:early_recomb}.  This effects mainly  introduce strong degeneracies among several parameters, while the fits to CMB, BAO and SNeIa can be kept well due to the existence of the curvature \cite{Sekiguchi:2020teg}.

\begin{figure}[h]
\centering
\includegraphics[width=15cm]{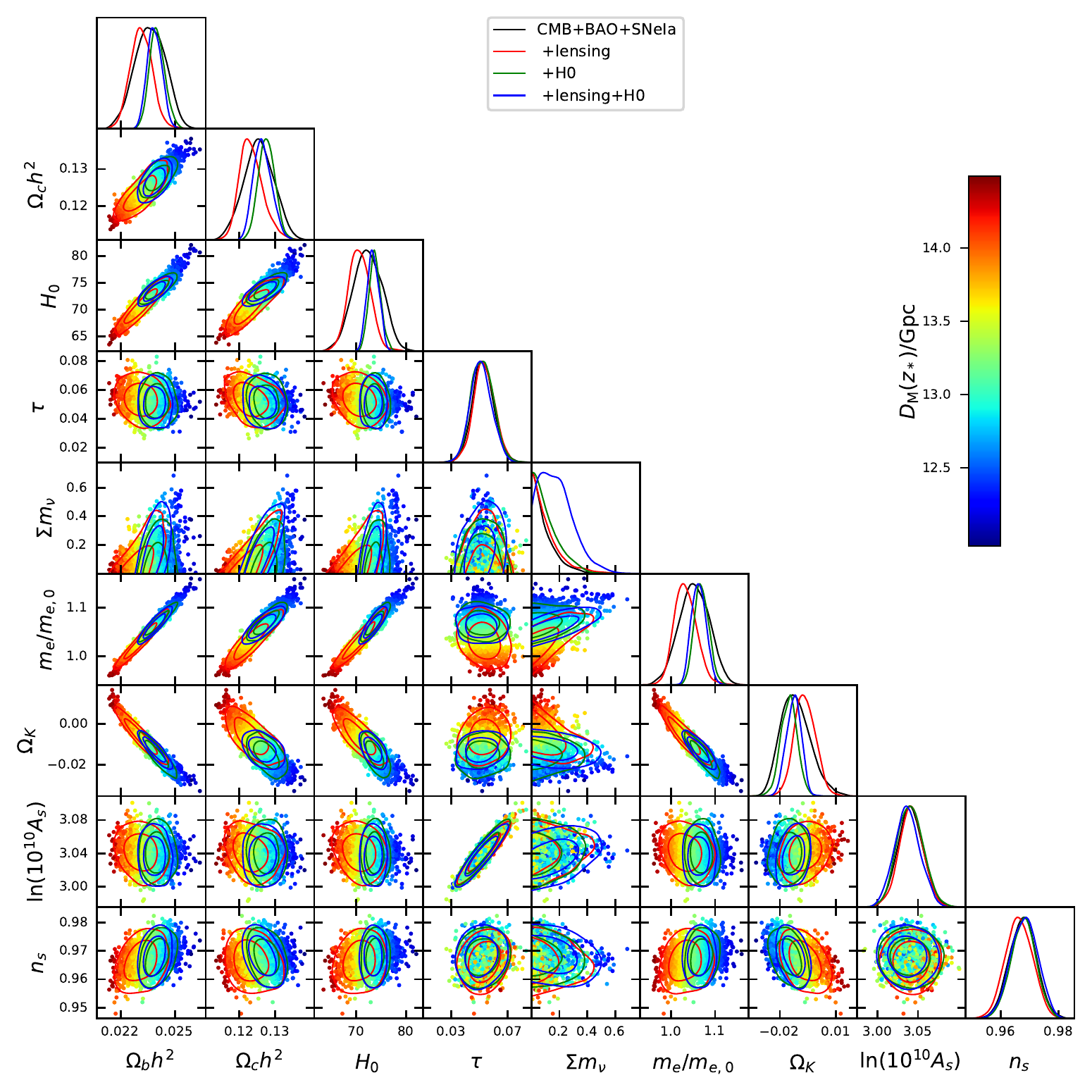}
\caption{Triangle plot of cosmological parameters in the $m_e \Omega_k  \Lambda$CDM model.  \label{fig:triangle_meOLCDM} }
\end{figure}

When $m_e$ is increased, the recombination epoch becomes earlier, which makes $r_s (z_\ast)$ smaller \cite{Hart:2017ndk}. To keep a good fit to CMB angular power spectra, we need to tune the acoustic scale $\theta_s (z_\ast) = r_s (z_\ast) / D_M(z_\ast)$ and hence $H_0$ can be increased  to cancel the effect due to the change of $m_e$, which introduces a strong degeneracy among $m_e$, $\Omega_bh^2$, $\Omega_ch^2$, $H_0$ and $\Omega_k$ as seen in Fig.~\ref{fig:triangle_meOLCDM}.

Interestingly, due to this severe degeneracy among several parameters in the $m_e\Omega_k\Lambda$CDM model, $\sum m_\nu$ and $H_0$ are now positively correlated and the direction of correlation follows a constant $D_M (z_\ast)$ line as can be observed in Fig.~\ref{fig:2D}. Because of this positive correlation, the bound on neutrino masses is pushed upward when the local $H_0$ measurement is included in the analysis. As shown in Table~\ref{tab:mnu}, the 95 \% upper bound on $\sum m_\nu$ is $0.28~{\rm eV}$ for the analysis of CMB+BAO+SNeIa, however, it becomes $0.31~{\rm eV}$ for  CMB+BAO+SNeIa+H0.

It should also be mentioned that the inclusion of the Planck lensing data tends to prefer a non-zero neutrino masses \cite{Aghanim:2018oex},  which can also make the upper bound on $\sum m_\nu$ weaker. When one includes the lensing and local $H_0$ measurement data in addition to CMB+BAO+SNeIa, the upper bound on neutrino masses $\sum m_\nu$ is 0.4~eV, which is fairly weak compared the counterpart in the $\Lambda$CDM framework.

As we have already emphasized, the  $m_e\Omega_k\Lambda$CDM model  shares the properties which should be satisfied by a successful model resolving the $H_0$ tension.  Therefore constraints on neutrino masses obtained in the framework of $m_e\Omega_k\Lambda$CDM model would show a general tendency in models where the $H_0$ tension is solved.

\section{Conclusion  \label{sec:conclusion}}

In this paper, we have investigated cosmological constraint on neutrino masses in the light of the $H_0$ tension. Since $H_0$ and neutrino masses $\sum m_\nu$ are correlated, particularly in CMB data, the $H_0$ tension would give a significant implications for cosmological bounds on neutrino masses. 

In the $\Lambda$CDM model, $H_0$ and neutrino masses are negatively correlated, which indicates that the upper bound on $\sum m_\nu$ becomes superficially tighter when the local $H_0$ measurement is included in the analysis as shown in Fig.~\ref{fig:1D} and Table~\ref{tab:mnu}.  However as seen from Fig.~\ref{fig:2D}, the value of $H_0$ indicated by CMB+BAO+SNeIa, even including other data sets,  is in large tension with the one obtained from the local $H_0$ measurement.  Therefore the cosmological bound on neutrino masses in the framework of $\Lambda$CDM cannot be taken at  face value if we take a position that the $H_0$ tension suggests the modification of the cosmological model.

In the light of this consideration, it would be indispensable to study the cosmological bound on neutrino masses in the framework where the $H_0$ tension is resolved. Although many models have been proposed for a solution to the $H_0$ tension, a model with early recombination in a non-flat Universe proposed in \cite{Sekiguchi:2020teg} satisfies the necessary conditions which a successful model should hold and hence we have investigated a cosmological constraint on neutrino masses in the framework of the $m_e \Omega_k\Lambda$CDM model. As emphasized, the analysis in this framework should give a general tendency in models where the $H_0$ tension can be solved.  To check how the assumption of a non-flat Universe affects a constraint on $\sum m_\nu$, we also made an analysis in the $\Omega_k\Lambda$CDM model as well.

From the analysis using the data of CMB+BAO+SNeIa, an upper bound on the neutrino masses in the $\Lambda$CDM,  $\Omega_k\Lambda$CDM and $m_e \Omega_k\Lambda$CDM models are $\sum m_\nu < 0.11~{\rm eV}, 0.16~{\rm eV}$ and $0.28~{\rm eV}$ (95 \% C.L.), respectively. In  the $\Omega_k\Lambda$CDM model, the curvature of the Universe $\Omega_k$ can change the angular diameter distance to last scattering surface, which generates a degeneracy between $\Omega_k$ and $\sum m_\nu$ and an upper bound on $\sum m_\nu$ gets weaker compare to the $\Lambda$CDM case.  When the varying $m_e$ is introduced (i.e., in the $m_e \Omega_k\Lambda$CDM model), the bound on $\sum m_\nu$ gets significantly weaker since the varying $m_e$ degenerates with several parameters \cite{Hart:2019dxi}. Furthermore, in the framework of $m_e \Omega_k\Lambda$CDM model, the correlation between $H_0$ and $\sum m_\nu$ is positive, differently from the case of $\Lambda$CDM model, and hence including the local $H_0$ measurement, the bound on $\sum m_\nu$ gets significantly looser. 

As argued in this paper, the $H_0$ tension can also affect other aspects of cosmology such as neutrino masses. We have investigated this issue and demonstrated that a cosmological bound on neutrino masses is actually affected in the light of the $H_0$ tension. Since the $H_0$ tension is now more than 5$\sigma$, we need to investigate further the implications of the $H_0$ tension to other aspects of cosmology.

\section*{Acknowledgements}
This work is supported by JSPS KAKENHI Grant Numbers 18H04339 (TS), 18K03640 (TS), 17H01131 (TT, TS), 19K03874 (TT) and
MEXT KAKENHI Grant Number 19H05110 (TT). This research was conducted using the Fujitsu PRIMERGY CX600M1/CX1640M1 
(Oakforest-PACS) in the Information Technology Center, The University of Tokyo.

\bibliography{neutrino_H0}

\providecommand{\href}[2]{#2}\begingroup\raggedright\begin{thebibliography}{10}

\bibitem{Zyla:2020zbs}
{\bf Particle Data Group} Collaboration, P.~Zyla et~al., {\it {Review of
  Particle Physics}},  {\em PTEP} {\bf 2020} (2020), no.~8 083C01.

\bibitem{Aker:2019uuj}
{\bf KATRIN} Collaboration, M.~Aker et~al., {\it {Improved Upper Limit on the
  Neutrino Mass from a Direct Kinematic Method by KATRIN}},  {\em Phys. Rev.
  Lett.} {\bf 123} (2019), no.~22 221802,
  [\href{http://arxiv.org/abs/1909.06048}{{\tt arXiv:1909.06048}}].

\bibitem{KamLAND-Zen:2016pfg}
{\bf KamLAND-Zen} Collaboration, A.~Gando et~al., {\it {Search for Majorana
  Neutrinos near the Inverted Mass Hierarchy Region with KamLAND-Zen}},  {\em
  Phys. Rev. Lett.} {\bf 117} (2016), no.~8 082503,
  [\href{http://arxiv.org/abs/1605.02889}{{\tt arXiv:1605.02889}}]. [Addendum:
  Phys.Rev.Lett. 117, 109903 (2016)].

\bibitem{Aghanim:2018eyx}
{\bf Planck} Collaboration, N.~Aghanim et~al., {\it {Planck 2018 results. VI.
  Cosmological parameters}},  \href{http://arxiv.org/abs/1807.06209}{{\tt
  arXiv:1807.06209}}.

\bibitem{Abbott:2017wau}
{\bf DES} Collaboration, T.~Abbott et~al., {\it {Dark Energy Survey year 1
  results: Cosmological constraints from galaxy clustering and weak lensing}},
  {\em Phys. Rev. D} {\bf 98} (2018), no.~4 043526,
  [\href{http://arxiv.org/abs/1708.01530}{{\tt arXiv:1708.01530}}].

\bibitem{Hikage:2018qbn}
{\bf HSC} Collaboration, C.~Hikage et~al., {\it {Cosmology from cosmic shear
  power spectra with Subaru Hyper Suprime-Cam first-year data}},  {\em Publ.
  Astron. Soc. Jap.} {\bf 71} (2019), no.~2 Publications of the Astronomical
  Society of Japan, Volume 71, Issue 2, April 2019, 43,
  https://doi.org/10.1093/pasj/psz010,
  [\href{http://arxiv.org/abs/1809.09148}{{\tt arXiv:1809.09148}}].

\bibitem{Cuesta:2015iho}
A.~J. Cuesta, V.~Niro, and L.~Verde, {\it {Neutrino mass limits: robust
  information from the power spectrum of galaxy surveys}},  {\em Phys. Dark
  Univ.} {\bf 13} (2016) 77--86, [\href{http://arxiv.org/abs/1511.05983}{{\tt
  arXiv:1511.05983}}].

\bibitem{Vagnozzi:2017ovm}
S.~Vagnozzi, E.~Giusarma, O.~Mena, K.~Freese, M.~Gerbino, S.~Ho, and
  M.~Lattanzi, {\it {Unveiling $\nu$ secrets with cosmological data: neutrino
  masses and mass hierarchy}},  {\em Phys. Rev. D} {\bf 96} (2017), no.~12
  123503, [\href{http://arxiv.org/abs/1701.08172}{{\tt arXiv:1701.08172}}].

\bibitem{Loureiro:2018pdz}
A.~Loureiro et~al., {\it {On The Upper Bound of Neutrino Masses from Combined
  Cosmological Observations and Particle Physics Experiments}},  {\em Phys.
  Rev. Lett.} {\bf 123} (2019), no.~8 081301,
  [\href{http://arxiv.org/abs/1811.02578}{{\tt arXiv:1811.02578}}].

\bibitem{Lattanzi:2017ubx}
M.~Lattanzi and M.~Gerbino, {\it {Status of neutrino properties and future
  prospects - Cosmological and astrophysical constraints}},  {\em Front. in
  Phys.} {\bf 5} (2018) 70, [\href{http://arxiv.org/abs/1712.07109}{{\tt
  arXiv:1712.07109}}].

\bibitem{Riess:2020sih}
A.~G. Riess, {\it {The Expansion of the Universe is Faster than Expected}},
  {\em Nature Rev. Phys.} {\bf 2} (2019), no.~1 10--12,
  [\href{http://arxiv.org/abs/2001.03624}{{\tt arXiv:2001.03624}}].

\bibitem{Fukugita:2006rm}
M.~Fukugita, K.~Ichikawa, M.~Kawasaki, and O.~Lahav, {\it {Limit on the
  Neutrino Mass from the WMAP Three Year Data}},  {\em Phys. Rev. D} {\bf 74}
  (2006) 027302, [\href{http://arxiv.org/abs/astro-ph/0605362}{{\tt
  astro-ph/0605362}}].

\bibitem{Sekiguchi:2009zs}
T.~Sekiguchi, K.~Ichikawa, T.~Takahashi, and L.~Greenhill, {\it {Neutrino mass
  from cosmology: Impact of high-accuracy measurement of the Hubble constant}},
   {\em JCAP} {\bf 03} (2010) 015, [\href{http://arxiv.org/abs/0911.0976}{{\tt
  arXiv:0911.0976}}].

\bibitem{Knox:2019rjx}
L.~Knox and M.~Millea, {\it {Hubble constant hunter's guide}},  {\em Phys. Rev.
  D} {\bf 101} (2020), no.~4 043533,
  [\href{http://arxiv.org/abs/1908.03663}{{\tt arXiv:1908.03663}}].

\bibitem{Joudaki:2012fx}
S.~Joudaki, {\it {Constraints on Neutrino Mass and Light Degrees of Freedom in
  Extended Cosmological Parameter Spaces}},  {\em Phys. Rev. D} {\bf 87} (2013)
  083523, [\href{http://arxiv.org/abs/1202.0005}{{\tt arXiv:1202.0005}}].

\bibitem{dePutter:2014hza}
R.~de~Putter, E.~V. Linder, and A.~Mishra, {\it {Inflationary Freedom and
  Cosmological Neutrino Constraints}},  {\em Phys. Rev. D} {\bf 89} (2014),
  no.~10 103502, [\href{http://arxiv.org/abs/1401.7022}{{\tt
  arXiv:1401.7022}}].

\bibitem{Chen:2016eyp}
Y.~Chen, B.~Ratra, M.~Biesiada, S.~Li, and Z.-H. Zhu, {\it {Constraints on
  non-flat cosmologies with massive neutrinos after Planck 2015}},  {\em
  Astrophys. J.} {\bf 829} (2016), no.~2 61,
  [\href{http://arxiv.org/abs/1603.07115}{{\tt arXiv:1603.07115}}].

\bibitem{DiValentino:2016ikp}
E.~Di~Valentino, S.~Gariazzo, M.~Gerbino, E.~Giusarma, and O.~Mena, {\it {Dark
  Radiation and Inflationary Freedom after Planck 2015}},  {\em Phys. Rev. D}
  {\bf 93} (2016), no.~8 083523, [\href{http://arxiv.org/abs/1601.07557}{{\tt
  arXiv:1601.07557}}].

\bibitem{Wang:2016tsz}
S.~Wang, Y.-F. Wang, D.-M. Xia, and X.~Zhang, {\it {Impacts of dark energy on
  weighing neutrinos: mass hierarchies considered}},  {\em Phys. Rev. D} {\bf
  94} (2016), no.~8 083519, [\href{http://arxiv.org/abs/1608.00672}{{\tt
  arXiv:1608.00672}}].

\bibitem{Canac:2016smv}
N.~Canac, G.~Aslanyan, K.~N. Abazajian, R.~Easther, and L.~C. Price, {\it
  {Testing for New Physics: Neutrinos and the Primordial Power Spectrum}},
  {\em JCAP} {\bf 09} (2016) 022, [\href{http://arxiv.org/abs/1606.03057}{{\tt
  arXiv:1606.03057}}].

\bibitem{Bellomo:2016xhl}
N.~Bellomo, E.~Bellini, B.~Hu, R.~Jimenez, C.~Pena-Garay, and L.~Verde, {\it
  {Hiding neutrino mass in modified gravity cosmologies}},  {\em JCAP} {\bf 02}
  (2017) 043, [\href{http://arxiv.org/abs/1612.02598}{{\tt arXiv:1612.02598}}].

\bibitem{Lorenz:2017fgo}
C.~S. Lorenz, E.~Calabrese, and D.~Alonso, {\it {Distinguishing between
  Neutrinos and time-varying Dark Energy through Cosmic Time}},  {\em Phys.
  Rev. D} {\bf 96} (2017), no.~4 043510,
  [\href{http://arxiv.org/abs/1706.00730}{{\tt arXiv:1706.00730}}].

\bibitem{Vagnozzi:2018jhn}
S.~Vagnozzi, S.~Dhawan, M.~Gerbino, K.~Freese, A.~Goobar, and O.~Mena, {\it
  {Constraints on the sum of the neutrino masses in dynamical dark energy
  models with $w(z) \geq -1$ are tighter than those obtained in $\Lambda$CDM}},
   {\em Phys. Rev. D} {\bf 98} (2018), no.~8 083501,
  [\href{http://arxiv.org/abs/1801.08553}{{\tt arXiv:1801.08553}}].

\bibitem{Zhao:2018fjj}
M.~Zhao, R.~Guo, D.~He, J.~Zhang, and X.~Zhang, {\it {Dark energy versus
  modified gravity: Impacts on measuring neutrino mass}},  {\em Sci. China
  Phys. Mech. Astron.} {\bf 63} (2020), no.~3 230412,
  [\href{http://arxiv.org/abs/1810.11658}{{\tt arXiv:1810.11658}}].

\bibitem{Poulin:2018zxs}
V.~Poulin, K.~K. Boddy, S.~Bird, and M.~Kamionkowski, {\it {Implications of an
  extended dark energy cosmology with massive neutrinos for cosmological
  tensions}},  {\em Phys. Rev. D} {\bf 97} (2018), no.~12 123504,
  [\href{http://arxiv.org/abs/1803.02474}{{\tt arXiv:1803.02474}}].

\bibitem{Zhang:2020mox}
M.~Zhang, J.-F. Zhang, and X.~Zhang, {\it {Impacts of dark energy on
  constraining neutrino mass after Planck 2018}},
  \href{http://arxiv.org/abs/2005.04647}{{\tt arXiv:2005.04647}}.

\bibitem{Sekiguchi:2020teg}
T.~Sekiguchi and T.~Takahashi, {\it {Early recombination as a solution to the
  $H_0$ tension}},  \href{http://arxiv.org/abs/2007.03381}{{\tt
  arXiv:2007.03381}}.

\bibitem{Hart:2017ndk}
L.~Hart and J.~Chluba, {\it {New constraints on time-dependent variations of
  fundamental constants using Planck data}},  {\em Mon. Not. Roy. Astron. Soc.}
  {\bf 474} (2018), no.~2 1850--1861,
  [\href{http://arxiv.org/abs/1705.03925}{{\tt arXiv:1705.03925}}].

\bibitem{Hart:2019dxi}
L.~Hart and J.~Chluba, {\it {Updated fundamental constant constraints from
  Planck 2018 data and possible relations to the Hubble tension}},  {\em Mon.
  Not. Roy. Astron. Soc.} {\bf 493} (2020), no.~3 3255--3263,
  [\href{http://arxiv.org/abs/1912.03986}{{\tt arXiv:1912.03986}}].

\bibitem{Aghanim:2019ame}
{\bf Planck} Collaboration, N.~Aghanim et~al., {\it {Planck 2018 results. V.
  CMB power spectra and likelihoods}},
  \href{http://arxiv.org/abs/1907.12875}{{\tt arXiv:1907.12875}}.

\bibitem{Beutler:2011hx}
F.~Beutler, C.~Blake, M.~Colless, D.~Jones, L.~Staveley-Smith, L.~Campbell,
  Q.~Parker, W.~Saunders, and F.~Watson, {\it {The 6dF Galaxy Survey: Baryon
  Acoustic Oscillations and the Local Hubble Constant}},  {\em Mon. Not. Roy.
  Astron. Soc.} {\bf 416} (2011) 3017--3032,
  [\href{http://arxiv.org/abs/1106.3366}{{\tt arXiv:1106.3366}}].

\bibitem{Ross:2014qpa}
A.~J. Ross, L.~Samushia, C.~Howlett, W.~J. Percival, A.~Burden, and M.~Manera,
  {\it {The clustering of the SDSS DR7 main Galaxy sample \textendash{} I. A 4
  per cent distance measure at $z = 0.15$}},  {\em Mon. Not. Roy. Astron. Soc.}
  {\bf 449} (2015), no.~1 835--847, [\href{http://arxiv.org/abs/1409.3242}{{\tt
  arXiv:1409.3242}}].

\bibitem{Alam:2016hwk}
{\bf BOSS} Collaboration, S.~Alam et~al., {\it {The clustering of galaxies in
  the completed SDSS-III Baryon Oscillation Spectroscopic Survey: cosmological
  analysis of the DR12 galaxy sample}},  {\em Mon. Not. Roy. Astron. Soc.} {\bf
  470} (2017), no.~3 2617--2652, [\href{http://arxiv.org/abs/1607.03155}{{\tt
  arXiv:1607.03155}}].

\bibitem{Scolnic:2017caz}
D.~Scolnic et~al., {\it {The Complete Light-curve Sample of Spectroscopically
  Confirmed SNe Ia from Pan-STARRS1 and Cosmological Constraints from the
  Combined Pantheon Sample}},  {\em Astrophys. J.} {\bf 859} (2018), no.~2 101,
  [\href{http://arxiv.org/abs/1710.00845}{{\tt arXiv:1710.00845}}].

\bibitem{Aghanim:2018oex}
{\bf Planck} Collaboration, N.~Aghanim et~al., {\it {Planck 2018 results. VIII.
  Gravitational lensing}},  {\em Astron. Astrophys.} {\bf 641} (2020) A8,
  [\href{http://arxiv.org/abs/1807.06210}{{\tt arXiv:1807.06210}}].

\bibitem{Lewis:2002ah}
A.~Lewis and S.~Bridle, {\it {Cosmological parameters from CMB and other data:
  A Monte Carlo approach}},  {\em Phys. Rev. D} {\bf 66} (2002) 103511,
  [\href{http://arxiv.org/abs/astro-ph/0205436}{{\tt astro-ph/0205436}}].

\bibitem{AliHaimoud:2010dx}
Y.~Ali-Haimoud and C.~M. Hirata, {\it {HyRec: A fast and highly accurate
  primordial hydrogen and helium recombination code}},  {\em Phys. Rev. D} {\bf
  83} (2011) 043513, [\href{http://arxiv.org/abs/1011.3758}{{\tt
  arXiv:1011.3758}}].

\bibitem{hyrec}
Y.~Ali-Haimoud, ``{Third release of HyRec (May 2012): technical explanatory
  supplement}.''
  \url{https://pages.jh.edu/~yalihai1/hyrec/supplement_may2012.pdf}.

\end{thebibliography}\endgroup

\end{document}